# Smart GSM Based Home Automation System

Rozita Teymourzadeh, *CEng, Member IEEE/IET*, Salah Addin Ahmed, Kok Wai Chan, and Mok Vee Hoong
Faculty of Engineering, Technology & Built Environment
UCSI University
Kuala Lumpur, Malaysia
rozita@ucsiuniversity.edu.my

*Abstract* — This research work investigates the potential of 'Full Home Control', which is the aim of the Home Automation Systems in near future. The analysis and implementation of the home automation technology using Global System for Mobile Communication (GSM) modem to control home appliances such as light, conditional system, and security system via Short Message Service (SMS) text messages is presented in this paper. The proposed research work is focused on functionality of the GSM protocol, which allows the user to control the target system away from residential using the frequency bandwidths. The concept of serial communication and AT-commands has been applied towards development of the smart GSM-based home automation system. Home owners will be able to receive feedback status of any home appliances under control whether switched on or off remotely from their mobile phones. PIC16F887 microcontroller with the integration of GSM provides the smart automated house system with the desired baud rate of 9600 bps. The proposed prototype of GSM based home automation system was implemented and tested with maximum of four loads and shows the accuracy of ≥98%.

## I. Introduction

In recent years, there has been a growing interest among consumers in the smart home concept [1]. Smart homes contain multiple, connected devices such as home entertainment consoles, security systems, lighting, access control systems and surveillance. Intelligent home automation system is incorporated into smart homes to provide comfort, convenience, and security to home owners [2-4]. Home automation system represents and reports the status of the connected devices in an intuitive, user-friendly interface allowing the user to interact and control various devices with the touch of a few buttons. Some of the major communication technologies used by today's home automation system [5-7] include Bluetooth, WiMAX and Wireless LAN (Wi-Fi), Zigbee, and Global System for Mobile Communication (GSM).

All GSM is one of the most widely used cellular technologies in the world [8,9]. With the increase in the number of GSM subscribers, research and development [10-12] is heavily supported in further investigating the GSM implementation. In 2009, Das, Sanaullah, et. al. [13] developed a cell phone based remote management and control system for home appliances. However, a few limitations for the system include not being able to control multiple appliances concurrently and the home automations system could not verify the status of the appliances. ElKamchouchi and ElShafee [6] presented the design and prototype implementation of basic home automation system based on SMS technology using AT89C55 Atmel microcontroller. The microcontroller acts as the bridge between the GSM network and sensors of the home automation system. Further researches have been conducted to analyze the performance of other home automation control system [14-16]. Internet and wireless communications have also been utilized in parallel with GSM for home automations [17].

Among the cellular technologies, GSM network is preferred for the communication between the home appliances and the user due to its wide spread coverage [8,9] which makes the whole system online for almost all the time. Another advantage of using the GSM network in home automation is its high security infrastructure, which provides maximum reliability whereby other people cannot monitor the information sent or received. Hence, this research work implements SMS based control for home appliances using the GSM architecture without accessing the local network.

## II. System Realization

The design of our proposed smart GSM-based home automation system is given in Fig. 1. The architecture consists of mobile phone and GSM modem. In the proposed system design, incoming SMS message is sent from the user phone to the GSM modem as a text message via cellular network. The GSM modem then sends the commands in text mode to the PIC microcontroller using an RS232 interface. The RS232 voltage levels are at ±12V whereas both the microcontroller input and output operates at 0V to +5V. Since RS232 is not compatible with microcontroller, MAX232 is utilized to enable the communication between both the GSM modem and PIC microcontroller by converting RS232 level signals to TTL level signal. Outgoing message from the system containing the home appliances status is delivered to the mobile phone through GSM modem.

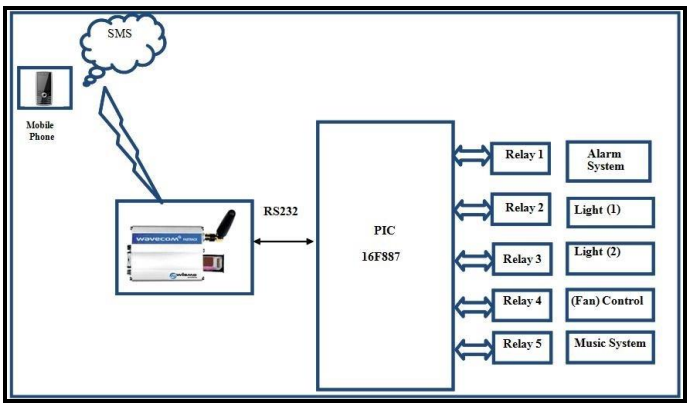

Fig. 1. Stage Realization.

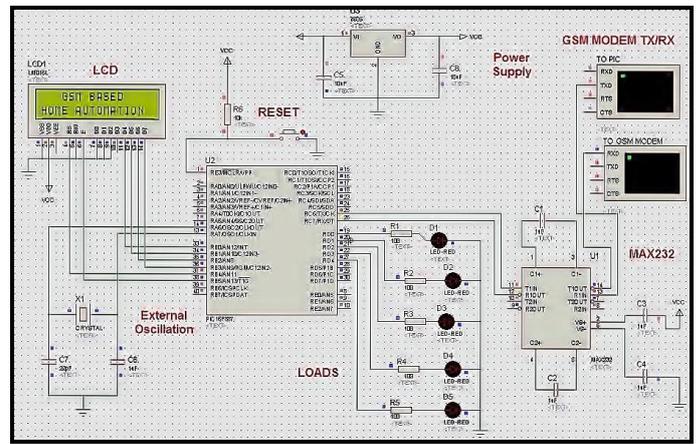

Fig. 2. Simulation of the proposed GSM based home automation system.

The 8-bit PIC16F887 microcontroller generally consists of timers, Analog to Digital Converters (ADCs), and Universal Synchronous Asynchronous Receiver Transmitter (USART). In the proposed research work, the microcontroller receives instructions and decodes them to give device address and command, then sends corresponding signals to the driver of the power circuit. In addition, the microcontroller ensures dual independent operation action to turn on the device or switch it off. A feedback status of any devices under control whether switched on or off will be provided by the microcontroller.

The RS-232 interface standard defines the electrical and mechanical details of the interface between Data Terminal Equipment (DTE) and Data Communications Equipment (DCE), which employ serial binary data interchange. The current version of the standard refers to DCE as Data Circuit terminating Equipment. Physically, interfacing between the PIC16F887 and GSM modem was implemented using RS232 standard installed on Max232.

Since the system design has not included a battery, an external power supply is connected to the system to drive sufficient amount of current through the circuit connections. Relays had been connected to the output loads for a stable electrical control because these relays can provide feeding for different voltage level loads. Hence, the output selection becomes easier at any voltage.

Fig. 2 demonstrates the simulation of the proposed GSM based home automation system being implemented in several stages. The PIC16F887 was simulated with the GSM modem by connecting it to the physical ports on the PC. MAX232 was placed to ensure proper transmission of data between the two terminals. Virtual terminal monitors the text sent from, and to the PIC16F887 while checking the transmission process and testing the algorithm. The waveform of the transmitted and received messages is monitored from the oscilloscope. While simulating the GSM modem, the "COMPIM" clock physically tests the response of the GSM modem by connecting it to a physical port on PC.

III. IMPLEMENTATION

As shown in Fig. 3, waveform is generated in the digital oscilloscope when the command is transmitted from PIC16F887 to GSM modem. This command will execute the deletion of the first message from the SIM card memory. The program reaches to a sleep state waiting for the new incoming text messages and then compares the text message with the stored commands. If both the received text message and the stored commands match, then, it will execute the intended command, which is turning on or off the output terminal.

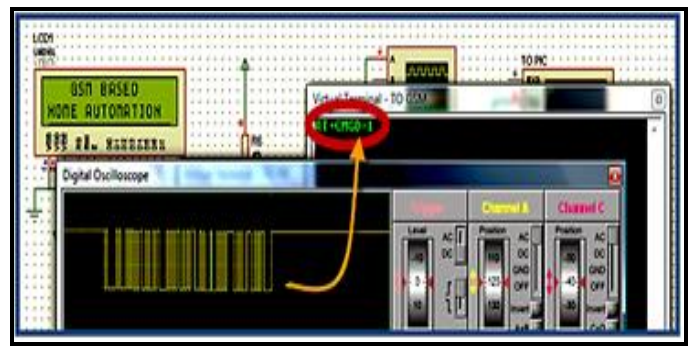

Fig. 3. **AT+CMGD=1** command to delete SMS message.

A simulation result was performed to observe the response while communicating with the GSM modem. The AT command was sent from PIC16F887 to the GSM modem as the program starts up, then the response is received from the GSM modem after very short period of time that does not exceed 500 microseconds which is fast enough to detect the incoming message from the modem. The display was used to show when the PIC is transmitting the signal.

From Fig. 4, the PIC16F887 serves as a transmitter initially and then, it receives the response from modem. The duration between the text sent and response received is less than 500 microseconds. Hence, the delay time is ignored because the data sent are limited to certain commands that do not exceed 4 or 5 characters. However, if the amount of data transferred is very huge, the response delay time is very critical and must be studied and analyzed in detail. Furthermore, the delay that must be controlled for this project is for GSM communication that is mostly governed by the SMS protocol and does not exceed 2 or 3 seconds as had been tested practically.

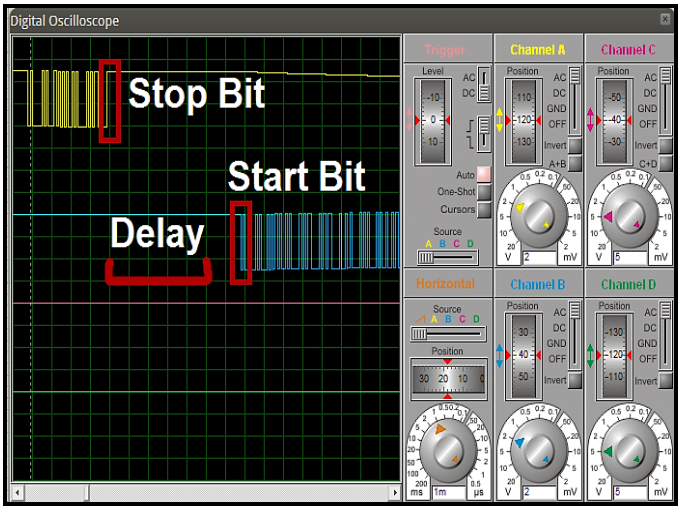

Fig. 4. Waveform of transceiver text message.

A layout is designed using Proteus ISIS Professional, while adding Virtual Terminal and connecting it to COMPIM, which is physically the DB9 connector in the PC terminal. It can be added from the library as shown in Fig. 5. The message alert text indicates that the message has been successfully received and stored in the SIM card in the memory location.

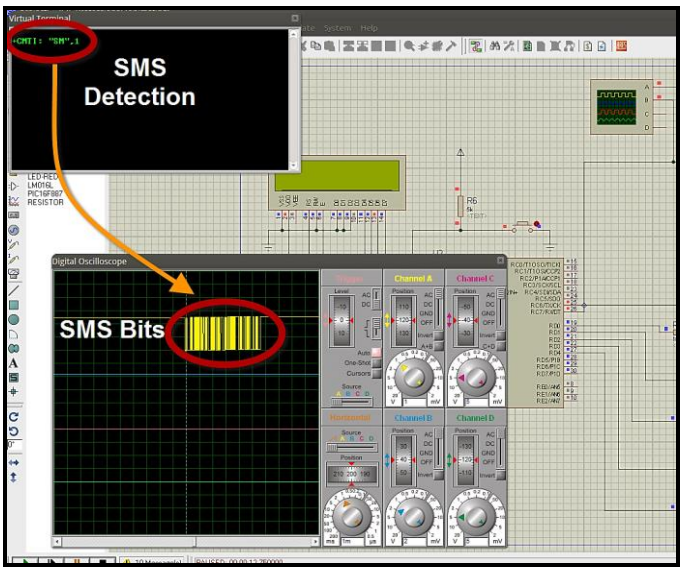

Fig. 5. Detection of incoming SMS from GSM modem.

The oscilloscope is connected to the receiver side (RX) on PIC16F887 microcontroller that is connected to GSM modem to display the waveform of the detected message as soon as it reaches the GSM memory. After detecting the message, the PIC16F887 is supposed to read and decode the incoming message to execute the intended command, which is turning on the loads. The whole process is being executed within short period of time that does not exceed 2 seconds. Fig. 6 shows the command text message had been sent to activate connected loads at the same time.

After receiving the command to turn on the lights, the relay set will perform switching operation from normally opened to close the circuit, allowing the current to start lighting as illustrated in Fig. 6.

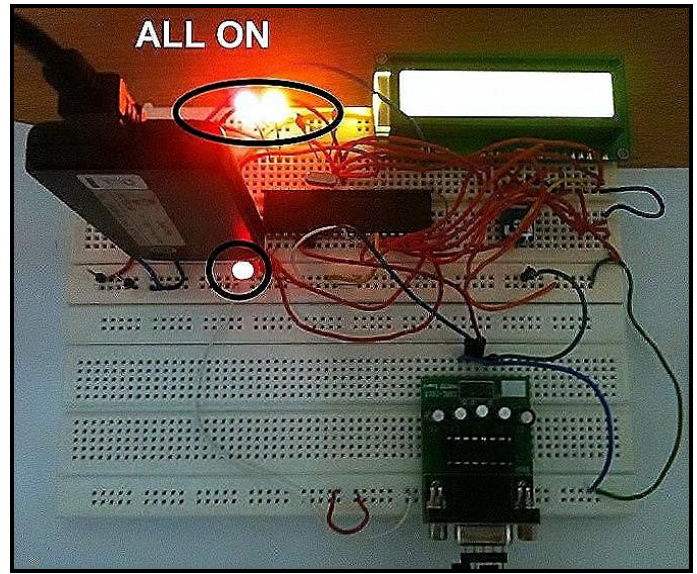

Fig. 6. All loads switched on when text commands sent via SMS.

## IV. RESULTS AND DISCUSSIONS

The Baud Rate Generator (BRG) is an 8-bit or 16-bit timer that is dedicated to the support of both asynchronous and synchronous EUSART operation. By default, the BRG operates in 8-bit mode. For the PIC16F887 with frequency of oscillator ($F_{OSC}$) of 20MHz, desired baud rate of 9600 bps, asynchronous mode, and 8-bit BRG, (1) will yield a baud rate of 9470 bps and an error of -1.35% illustrated in Fig. 7.

$$Calculated\ Baud\ Rate = \frac{FOSC}{64([SPBRGH:SPBRG]+1)} \quad (1)$$

| BAUD RATE | SYNC = 0, BRGH = 0, BRG16 = 0 | | | | | | | | | | | |
|---|---|---|---|---|---|---|---|---|---|---|---|---|
| | Fosc = 20.000 MHz | | | Fosc = 18.432 MHz | | | Fosc = 11.059 MHz | | | Fosc = 8.000 MHz | | |
| | Actual Rate | % Error | SPBRG value (decimal) | Actual Rate | % Error | SPBRG value (decimal) | Actual Rate | % Error | SPBRG value (decimal) | Actual Rate | % Error | SPBRG value (decimal) |
| 300 | — | — | — | — | — | — | — | — | — | — | — | — |
| 1200 | 1221 | 1.73 | 255 | 1200 | 0 | 239 | 1200 | 0 | 143 | 1202 | 0.16 | 103 |
| 2400 | 2404 | 0.16 | 129 | 2400 | 0 | 119 | 2400 | 0 | 71 | 2404 | 0.16 | 51 |
| 9600 | 9470 | -1.36 | 32 | 9600 | 0 | 29 | 9600 | 0 | 17 | 9615 | 0.16 | 12 |
| 10417 | 10417 | 0 | 29 | 10286 | -1.26 | 27 | 10165 | -2.42 | 16 | 10417 | 0 | 11 |
| 19.2k | 19.53k | 1.73 | 15 | 19.20k | 0 | 14 | 19.20k | 0 | 8 | — | — | — |
| 57.6k | — | — | — | 57.60k | 0 | 7 | 57.60k | 0 | 2 | — | — | — |
| 115.2k | — | — | — | — | — | — | — | — | — | — | — | — |

Fig. 7. Baud rate specification.

The prototype of the proposed GSM based home automation system is shown in Fig. 8. A 12V is supplied to the voltage regulator to power the circuit. MAX232 is connected to the GSM modem and the system run smoothly after the text message was received. The output loads were activated and automated the house in desirable basis.

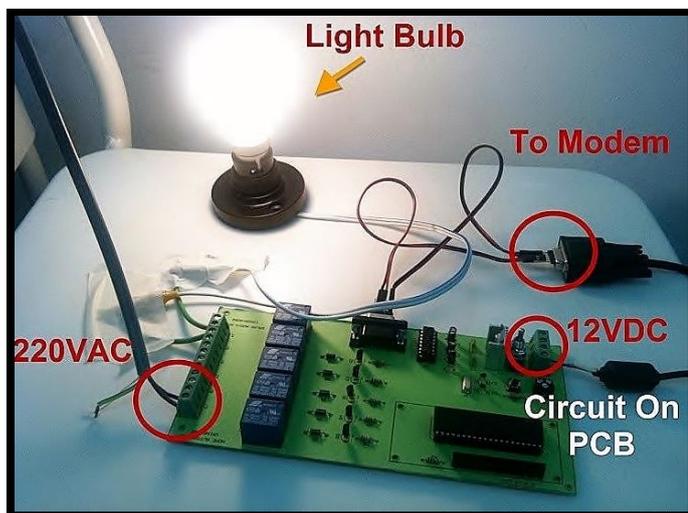

Fig. 8. Prototype of proposed GSM based home automation system.

V. CONCLUSION

Recently, the home automation market is very promising field that is growing very fast and needs vast range of developments that can be carried out in the concept of smart home. In this project design and implementation of smart GSM house was considered. PIC16F887 microcontroller with the cooperation of GSM provides the smart automated house system with the desired baud rate of 9600 bps. The proposed prototype was implemented and tested with maximum of four loads and shows the accuracy of ≥98%.